\documentclass[twocolumn,showpacs,preprintnumbers,amsmath,amssymb]{revtex4}

\usepackage{graphicx}
\usepackage{dcolumn}
\usepackage{bm}

\begin{document}

\title{Laser wakefield acceleration by petawatt ultra-short laser pulses }
\author{L. M. Gorbunov}\email{
gorbun@sci.lebedev.ru} \affiliation{P. N. Lebedev Physics
Institute, Russian Academy of Sciences, Leninskii prospect 53,
Moscow 119991, Russia}
\author{S. Yu.\ Kalmykov}
   \affiliation{ The Department of Physics and Institute for
Fusion Studies, The University of Texas at Austin, One University
Station C1500, Austin, Texas 78712}\affiliation{Centre de Physique
Th\'eorique (UMR 7644 du CNRS), \'Ecole Polytechnique, 91128 Palaiseau,
France}\author{P. Mora}\affiliation{Centre de Physique
Th\'eorique (UMR 7644 du CNRS), \'Ecole Polytechnique, 91128 Palaiseau,
France}
\date{\today}

\begin{abstract}
An ultra-short (about 30 fs) petawatt laser pulse focused with a
wide focal spot (about $100$~$\mu$m) in a rarefied plasma
($n_0\sim10^{17}$cm$^{-3}$) excites a nonlinear plasma wakefield
which can accelerate injected electrons up to the GeV  energy
without any pulse channelling. In these conditions, propagation of
the laser pulse with an over-critical power for relativistic
self-focusing is almost the same as in vacuum. The nonlinear
quasi-plane wake plasma wave, whose amplitude and phase velocity
vary along the laser path, effectively traps and accelerates
injected electrons with a wide range of initial energies.
Electrons accelerated along two Rayleigh lengths (about eight
centimeters) can gain an energy up to 1 GeV. In particular, the
electrons trapped from quite a long ($\tau_b\sim330$~fs)
non-resonant electron beamlet of 1 MeV particles eventually form a
low emittance bunch with energies in the range $900\pm50$ MeV. All
these conclusions follow from two-dimensional simulations
performed in cylindrical geometry by means of the fully
relativistic time-averaged particle code WAKE.
\end{abstract}

\pacs{52.35 Mw, 52.38 Hb, 52.38 Kd}

\maketitle

\section{\label{Sec1}Introduction}
In the original, or "standard", scheme of laser wakefield
acceleration (LWFA)~\cite{1,2}, a laser pulse of duration smaller
than a period of plasma oscillation $\tau_p\equiv2\pi/\omega_p$
excites a wake electron plasma wave (wakefield) with a phase
velocity close to the speed of light [here and elsewhere,
$\omega_p=(4\pi n_0e^2/m)^{1/2}$ is the electron plasma frequency,
$n_0$ is the electron plasma density, $m$ and $-|e|$ are the
electron mass at rest and charge].  When the electron density
perturbation exceeds the background density, the accelerating
electric field of the nonlinear wake can reach tens of GV/m, which
is by three orders of magnitude higher than that can be achieved
in the conventional accelerators without material breakdown. To
excite such a wakefield, a short laser pulse of very high
intensity is needed, that is, $I_0\lambda_0^2>1.4\times10^{18}$
Wcm$^{-2}\mu$m$^2$ (where $\lambda_0$ is a laser wavelength).
Reaching this range of intensities for lower laser energy in
previous experiments~\cite{3} made necessary a tight beam
focusing. The laser focal spot with a radius $r_0\sim10$~$\mu$m
gave a longitudinal extent of accelerating plasma (estimated as
two Rayleigh lengths $2Z_R=2\pi r_0^2/\lambda_0$) of the order of
a few millimeters. Hence, the observed energy gain of electrons
injected in the wake was moderate in spite of a high accelerating
gradient. Also, very narrow plasma wake driven by a tightly
focused laser is ineffective for trapping and acceleration of
injected electrons: the scale of radial variation was considerably
smaller than the axial wavelength, and most of the electrons were
expelled by strong radial electric fields. Only few electrons were
trapped and accelerated in the three-dimensional  potential
wells~\cite{3}.

The petawatt ($10^{15}$~W) CPA lasers of new generation, which are
under construction now~\cite{Aoyama}, are capable of delivering
ultra-short pulses (tens of femtoseconds) which can be focused
with a relatively large focal spot ($r_0\sim 100$~$\mu$m) yet
having a relativistic intensity ($I_0>10^{18}$~W/cm$^2$) on axis.
In this case, the effective acceleration distance (two Rayleigh
lengths) is naturally elongated up to about ten centimeters, and
the GeV energy range can be achieved by using the standard LWFA
technique without external optical guiding~\cite{2}. In the scheme
proposed, the plasma wavelength is smaller than the wake radial
size. The large transverse extent of the wakefield structure
reduces the detrimental effect of radial forces on accelerated
electrons.

Anticipating this attractive chance to reach the GeV range of
electron energy by using the standard LWFA,  we analyze in this
paper theoretically and numerically the details of laser pulse and
wakefield evolution, and simulate the acceleration of an electron
bunch in a wide range of parameters typical of those prospective
experiments.  The simulations have been carried out using the code
WAKE~\cite{4,5}. The laser parameters in the simulations are
fixed: the laser wavelength $\lambda_0=0.8$~$\mu$m, the pulse
energy 30~J, the pulse duration $\tau_L=30$~fs, the radius of
focal spot $r_0=100$~$\mu$m (hence, the maximum intensity in
vacuum is $I_0\approx6.4\times10^{18}$~W/cm$^2$). In the
simulations of electron acceleration the electron density of
unperturbed plasma is $n_0=1.12\times10^{17}$~cm$^3$, which gives
the period of plasma oscillation $\tau_p\approx330$~fs and the
plasma wavelength $\lambda_p=2\pi/k_p\approx100$~$\mu$m (where
$k_p=\omega_p/c$). Then, the normalized pulse duration and width
are $\omega_p\tau_L=0.56$ and $k_pr_0=6.28$, respectively. The
relativistic factor corresponding to the laser group velocity
$\gamma_g\equiv\omega_0/\omega_p=125$. The laser power is exceeds
by a factor of four the critical power for the relativistic
self-focusing $P_{cr}=16.2(\omega_0/\omega_p)^2$~GW~\cite{6}.

The paper is organized as follows.  Numerical study of nonlinear
effects in propagation of ultra-short petawatt laser pulse through
rarefied plasma is given in Sec.~\ref{Sec2}. The simulations
supported by  analytical considerations show that, for the
parameters chosen, mutual cancellation of the relativistic and
ponderomotive nonlinearities occurs in the pulse body. Thereby,
only for very high intensities and/or electron densities (such
that the critical power $P_{cr}$ is exceeded by more than a factor
of four) the nonlinearity produces some effect on the pulse shape.

Section~\ref{Sec3} addresses the effect of laser pulse transverse
evolution on the phase velocity of the two-dimensional (2D)
nonlinear wakefield. In our framework, the effect originates from
the relativistic nonlinearity of the plasma wake. The amplitude of
the wakefield is proportional to the laser intensity which grows
(drops) while the pulse focuses (diverges). On the other hand, the
relativistic reduction in frequency is proportional, in the weakly
relativistic approximation, to the wake amplitude squared.
Therefore, as the pulse approaches the focal plane, the wake
period stretches and its phase velocity drops below the pulse
group velocity $v_g=c^2k_0/\omega_0$. Beyond the focal plane the
wake period shrinks and tends at infinity to the linear limit
$\tau_p$. At this stage, the wake phase accelerates and can become
superluminous (i.e., exceeds not only the pulse group velocity in
plasma but the vacuum speed of light). A similar effect was
discussed previously for the case of longitudinally inhomogeneous
plasmas \cite{7}. Growth of the wake amplitude combined with the
reduction in phase velocity provides conditions for highly
efficient trapping of non-resonant injected electrons,
$\gamma_{e0}\ll\gamma_g$, in the stage of laser focusing [here and
below, $\gamma_{e0}\equiv(1-v_{e0}^2/c^2)^{-1/2}$ is the initial
Lorentz factor of the injected electrons corresponding to the
velocity $v_{e0}$ at the point of injection].

Wakefield acceleration of  resonant electron bunches is considered
in Sec.~\ref{Sec4}. Numerical simulations prove that the resonant
conditions $\gamma_{e0}\approx\gamma_g$ or
$\gamma_{e0}\approx\gamma_{p0}$ (where $\gamma_{p0}$ is a
relativistic factor given by the local phase velocity of wake at
the point of electron injection) are beneficial for the
ultra-short electron bunches only that are loaded directly in the
accelerating and focusing quarter of the wake period
($\tau_p/4\approx80$ fs for the parameters given). Taking a longer
bunch leads to a considerable energy spread because the resonant
electrons injected at the bottom of the potential well stay there
all the time and gain just a little energy. Injection of
mono-energetic electron bunches with $\gamma_{e0}=\gamma_g=125$ or
   $\gamma_{e0}=42$ in the second period of wakefield gives very
similar output: after two Rayleigh lengths (about eight centimeters)
the trapped electrons possess a broad energy spectrum
(about 100\% spread) with maxima at 0.5 GeV (for $\gamma_{e0}=125$)
or 0.75 GeV (for $\gamma_{e0}=42$) and a cutoff at approximately
1~GeV.  Tuning the energy of injected electrons to the resonance
with a given period of wakefield reduces the final rms emittance
of the accelerated electron bunch but has no effect on the final
energy spread.

Acceleration of electron bunches injected with energies far below
resonant is discussed in Sec.~\ref{Sec5}. Simulations show that
the electrons with initial energies 5 MeV and 1 MeV are not only
effectively trapped and accelerated up to 1~GeV, but also reveal
substantially less energy spread and lower root-mean-square (rms)
emittance than in the resonant case discussed in Sec.~\ref{Sec4}.
The slow electrons ($\gamma_{e_0}=0.08\gamma_g$ and
$\gamma_{e_0}=0.016\gamma_g$) loaded near the bottom of the 2D
potential well slip into the accelerating and focusing phase of
the wake period and get effectively accelerated. In particular,
acceleration of the 1~MeV electrons along twice the Rayleigh
length  produces a group of electrons with the energy $900\pm50$
MeV and rms emittance less than $10^{-4}$~mm~mrad.

Section~\ref{Sec6} discusses and summarizes the results obtained.
In Appendix~\ref{AppA} the analytical considerations are given of
those nonlinear phenomena which have an effect on the propagation
of ultra-short laser pulses in plasmas. The features of the plasma
wakefield generated by diffracting Gaussian short laser beam are
considered in Appendix~\ref{AppB}.

\section{\label{Sec2}Short laser pulse propagation in tenuous plasma:
compensation of relativistic and ponderomotive nonlinearities}

Relativistic and ponderomotive nonlinearities of a short
($\omega_p\tau_L<1$) and broad ($k_pr_0>1$) over-critical laser
pulse ($P>P_{cr}$) partly cancel each other~\cite{8}. The
vacuum-like propagation of such pulse~\cite{9}, however, is not
immune to the residual effect of the non-compensated relativistic
nonlinearity (see Appendix~\ref{AppA}). We simulate the laser
propagation, plasma wakefield excitation, and acceleration of
injected test electrons by means of the 2D axially symmetric fully
relativistic time-averaged particle code WAKE~\cite{4,5}, which is
based on the quasi-static~\cite{8} and extended paraxial~\cite{10}
approximations. The normalized slowly varying amplitude (envelope)
of laser vector potential $a(\xi,r,z)$ is evaluated as a function
of radial ($r$) and longitudinal ($z$) space variables and the
retarded time $\xi/c=t-z/c$. In the simulation, the pulse
propagates from the left to the right through a plasma slab
centered at the vacuum focal plane $z=0$. Simulation starts at a
plane $z=z_0<0$ and terminates at a symmetric plane $z=|z_0|$.
Initial laser beam with the parameters listed in Introduction is
Gaussian~\cite{12},
\begin{equation}
\label{1}
a=a_0(r,\xi,\zeta_0)\exp\left[-2\ln2(\xi-\xi_0)^2/(c\tau_L)^2 +
i\Psi_0\right],
\end{equation}
where $\zeta_0=-z_0/Z_R$, $ \Psi_0
=(r/r_0)^2\zeta_0/(1+\zeta_0^2)-\arctan \zeta_0$; $a_0(r,\zeta_0)
= a_0(1+\zeta_0^2)^{-1/2}\exp[-(r/r_0)^2/(1+\zeta_0^2)]$;
$\xi=\xi_0$ corresponds to the pulse center;  and $\tau_L$ is the
pulse full width at half-maximum (FWHM) in intensity. At the
starting point, the laser pulse (\ref{1}) has a converging
(concave) phase front and focuses into plasma. The simulation
proceeds from $z=-Z_R$ to $z=Z_R$, where $Z_R\approx4$~cm  for the
parameters chosen. Normalized amplitude of laser in the vacuum
focal spot is $a_0=1.72$.

\begin{figure}[t]
\includegraphics[scale=1]{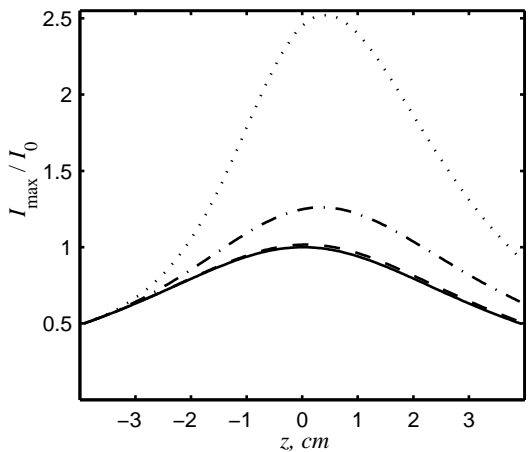}
\caption{\label{Fig1}The maximum intensity of laser pulse on the
axis as a function of propagation distance. The vacuum focal plane
is $z=0$. The intensity is normalized to the maximum intensity of
laser pulse in the  focal plane in vacuum,
$I_0=6.4\times10^{18}$~W/cm$^2$. Solid line corresponds to
focusing in vacuum; other lines correspond to focusing into
plasmas: dashed line --- $n_0/n_c=1.6\times10^{-5}$ ($P=P_{cr}$),
dash-dotted line --- $n_0/n_c=6.4\times10^{-5}$ ($P=4P_{cr})$,
dotted line --- $n_0/n_c=1.28\times10^{-4}$ ($P=8P_{cr})$.
   }
\end{figure}

A few simulations with different plasma densities demonstrate the
effect of nonlinearities on the pulse evolution. The maximum laser
intensity on the axis ($r=0$) is plotted in Fig.~\ref{Fig1} versus
propagation distance. The density
$n_0=n_{SF}=2.8\times10^{16}$~cm$^{-3}$ gives $P=P_{cr}$ (dashed
line). At $n_0=8n_{SF}$ (or $P=8P_{cr}$, dotted line), the
non-compensated relativistic self-focusing increases the intensity
by a factor of 2.5 in the vicinity of the focal plane. However,
for $n_0=4n_{SF}$ (or $P=4P_{cr}$), the on-axis laser intensity
varies similarly to the vacuum case (dash-dotted line in
Fig.~\ref{Fig1}). Therefore, under appropriate choice of
parameters, propagation of a relativistically strong ultra-short
overcritical laser pulse in a rarefied plasma is almost unaffected
by the relativistic self-focusing, which could be a challenge for
a longer pulse~\cite{6}.

\begin{figure}[t]
\includegraphics[scale=1]{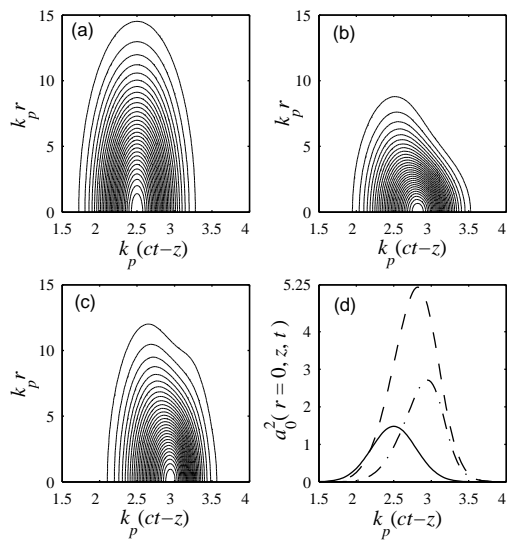}
\caption{\label{Fig2}Contour plots of normalized pulse intensity
at the three positions: $z=-Z_R$ (a), $z=0$ (b), $z=Z_R$ (c). At
$z=-Z_R$ the  pulse is Gaussian~(\ref{1}) with the amplitude
$a_0=1.72$, radius $k_pr_0=7.84$, and duration
$\omega_p\tau_L=0.68$, which give $P\approx6P_{cr}$. The solid,
dashed and dotted-dashed lines in the plot (d) show the on-axis
profile of intensity for the plots (a), (b), and (c).
   }
\end{figure}

Figure~\ref{Fig2} demonstrates the distortion of the laser radial
and temporal profiles due to the decompensation of relativistic
self-focusing. For $P\approx6P_{cr}$, the contour plots of
normalized intensity $a^2(\xi,r)$ show that the pulse leading part
spreads in the course of propagation, while the trailing part
shrinks. The effect manifests in full in the vicinity of the
vacuum focal plane. The difference between the speed of light in
vacuum and the pulse group velocity in plasma brings about the
gradually growing shift of the pulse center from its initial
position in the co-moving variables.

The analytical consideration given in a weakly relativistic
approximation in Appendix~\ref{AppA} predicts an amplitude
threshold $a_{0c}$ [see Eq.~(\ref{A12})], below which the laser
pulse radially spreads according to the linear theory of
diffraction for Gaussian beams~\cite{12}.  For a Gaussian temporal
profile, the critical amplitude reads $ a_{0c}=
(\omega_p\tau_L)^{-1}\sqrt{4\ln 2/[1+(k_pr_0/4)^2]}$. The laser
amplitude $a_0=1.72$ taken in the simulation of Fig.~\ref{Fig2}
exceeds the critical amplitude $a_{0c}\approx1.11$, which explains
the obvious manifestation of the nonlinearity in Fig.~\ref{Fig2}.
In order to reduce the adverse effect of decompensated
nonlinearities, the wakefield evolution and electron acceleration
will be further simulated in more rarefied plasmas. The parameters
of the case $P=4P_{cr}$ will be taken, that is, the plasma density
$n_0=1.12\times10^{17}$~cm$^{-3}$ that gives $a_{0c}\approx1.6$.
The relativistic factor of the laser pulse then amounts to
$\gamma_g=125$, and the normalized pulse duration to
$\omega_p\tau_L\approx0.56$.

\section{\label{Sec3}Excitation of nonlinear plasma wakefield
by short diffracting nearly Gaussian laser pulse}

\begin{figure}[t]
\includegraphics[scale=1]{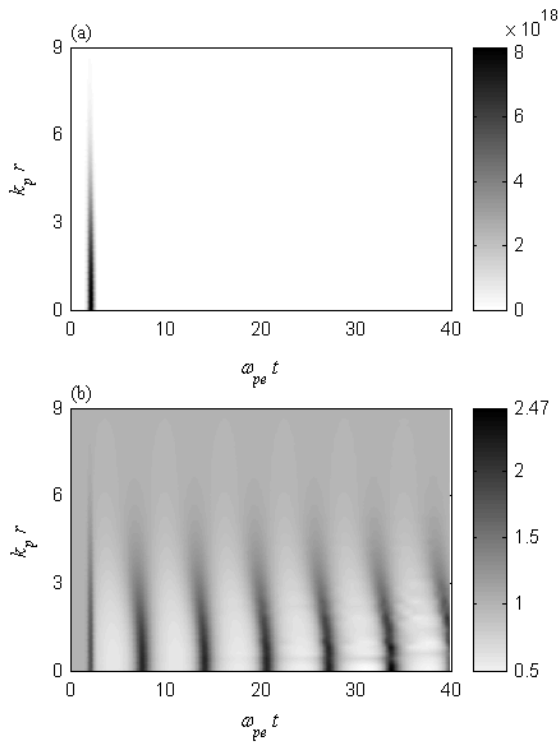}
\caption{\label{Fig3}Temporal evolution of the radial profile of
laser pulse intensity (in W/cm$^2$) (a) and  electron density
normalized to the background density  $n_0$ (b) in the vacuum
focal plane $z=0$.
   }
\end{figure}

We study numerically and analytically the plasma wakefield
evolution in the regime with laser nonlinearities mostly
compensated and the laser beam close to Gaussian; the parameters
are taken as suggested at the end of section~\ref{Sec2}. Although
the laser pulse is much shorter than a plasma period, the
intensity on axis is high enough to produce in the wake an
electron density perturbation of the order of the background
plasma density. The wake becomes strongly nonlinear in the
vicinity of focal plane, $z=0$, where its amplitude is maximal.
The radial and temporal profiles of the laser pulse and electron
density at this plane are shown in Fig.~\ref{Fig3}. The wakefield
is far from harmonic: the regions of density depression are much
wider than the density humps. Moreover, the wake phase front is
not plane and its curvature builds up with time. The relativistic
shift of plasma frequency in the wake is proportional, to the
lowest order, to the laser intensity squared, which radially
varies and thus brings about the said curvature~\cite{13}. The
radial phase variation, however, does not produce a radial
wavebreaking~\cite{14} within at least six wake periods.

The wake phase velocity is not constant along the laser path. This
effect has a purely relativistic origin. As the pulse focuses at
$z<0$, its intensity grows, and so does the relativistic shift of
plasma frequency. The wake period stretches, and the phase
velocity drops below the group velocity of laser. Beyond the focal
plane, $z>0$, the pulse radially spreads, and the plasma period
gradually shrinks thus tending back to the linear limit $\tau_p$.
At this stage, the wake phase velocity may exceed both laser pulse
group velocity in plasma and the speed of light in vacuum.
\begin{figure}[t]
\includegraphics[scale=1]{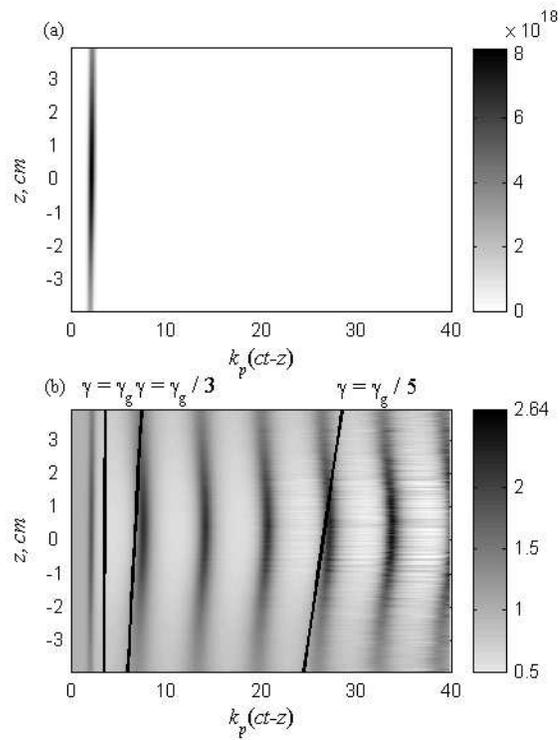}
\caption{\label{Fig4}On-axis evolution of laser intensity (in
W/cm$^2$) (a) and  electron density normalized to the background
density $n_0$ (b) as a function of retarded time and laser pulse
propagation length in the simulation of Fig.~\ref{Fig3}. The solid
lines in the plot (b) are the tangents to the electron density
crests. They characterize the local phase velocity of the wake at
$z=z_0=-Z_R$. Corresponding relativistic factors in units of
$\gamma_g=125$ are shown at the top line of the plot (b).
   }
\end{figure} The luminous point, where the wake phase reaches
the vacuum speed of light, does can exist on the path of a laser
pulse in a longitudinally inhomogeneous plasma~\cite{7}. In our
framework, it is due to the nonlinear frequency variation of
plasma wake driven by the diffracting laser. Figure~\ref{Fig4}
shows the on-axis laser intensity (a) and the electron density (b)
versus retarded time and distance $z$ in plasma. The plot (b)
demonstrates the phase ``deceleration'' (``acceleration'') at
$z<0$ ($z>0$) with the wake period stretching (contracting). At
$z>0$ the wake phase is superluminous.

An electron falls in resonance with the accelerating wakefield if
its velocity $v_{e0}$ coincides at the point of injection  with
the local phase velocity of wake. Taking tangents to the electron
density crests in Fig.~\ref{Fig4}(b) helps to evaluate the
resonant Lorentz factor $\gamma_{e0}\gg1$ of the electron injected
on axis at $z_0=-4$~cm. The tangent equation,
$z\approx2\xi\gamma_{e0}^2$,  gives
$\gamma_{e0}\approx42\approx\gamma_g/3$ for the second, and
$\gamma_{e0}\approx21=\gamma_g/5$ for the fifth wake period.
Weakly nonlinear theory of the wakefield excitation by a
diffracting short Gaussian laser pulse (see Appendix~\ref{AppB})
expresses tThe wake Lorentz factor as a function of normalized
pulse position $\zeta=z/Z_R$ and time delay $\xi/c$ [see
Eq.~(\ref{B8})], which gives $\gamma_{e0}$ by a factor of two
higher than inferred from the graphical estimates of
Fig.~\ref{Fig4}(b).

\section{\label{Sec4} Acceleration of resonant electron bunches}

\begin{figure}[t]
\includegraphics[scale=1]{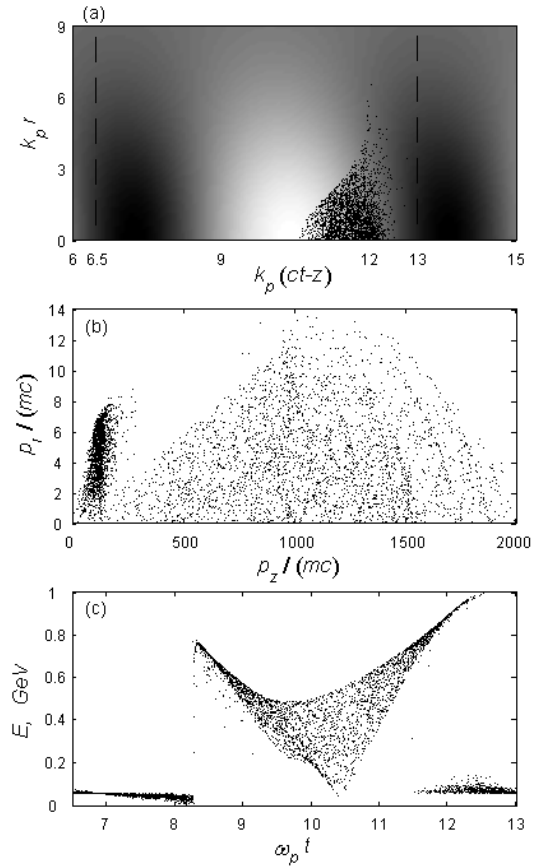}
\caption{\label{Fig5}Acceleration of the resonant electron bunch
($\gamma_{e0}=\gamma_g=125$). The space distribution (a), momentum
distribution (b), and energy versus the injection phase (c) are
shown for test electrons crossing the extraction plane $z=Z_R$.
Each dot corresponds to a numerical electron. The density of dots
characterizes the number of test particles per elementary volume
$k_p^2drd\xi$ in the plot (a) and per elementary volume of
momentum space $(dp_rdp_z)/(mc)^2$ in the plot (b). The grayscale
density in the plot (a) is proportional to the normalized
wakefield potential $\langle\psi\rangle$, the lightest gray
corresponds to $\langle\psi\rangle_{\max}=0.3$, the darkest gray
--- to $\langle\psi\rangle_{\min}=-0.24$.  Dashed
lines in the plot (a) show the boundaries of the bunch at the
injection point.
   }
\end{figure}
The test electron bunch is injected in the laser wake at
$z_0=-Z_R$. At this point, a Monte-Carlo generator creates
mono-energetic particles with the energy $mc^2\gamma_{0e}$ and
uniform distribution over time interval $\tau_b$. The bunch has a
transverse momentum spread, which gives a nonzero initial angular
divergence $\langle\alpha^2\rangle^{1/2}=\alpha_e$, and a nonzero
initial rms emittance $\varepsilon_\perp=(1/2)[\langle
r_\perp^2\rangle \langle (p_\perp/p)^2\rangle-\langle({\bf
r}_\perp \cdot {\bf p}_\perp)/p\rangle^2]^{1/2}$
(Ref.~\onlinecite{15}). Radial distribution of electron density in
the bunch is Gaussian with an rms radius $\sigma\sim r_0$. The
accelerated electrons are extracted from the wake at $z=Z_R$.

First,the conventional resonant condition $\gamma_{e0}=\gamma_g$
is considered. A bunch of 5000 test electrons with
$\gamma_{e0}=125$, zero angular divergence, and
$k_p\sigma=k_pr_0/\sqrt{2}=4.5$ is accelerated in the second wake
period ($6.5\le k_p\xi \le13$, which corresponds to the bunch
duration $\tau_b\approx330$~fs). Distributions of test electrons
in coordinate and momentum space at $z=Z_R$ , and final energy
versus injection phase are shown in Fig.~\ref{Fig5}. The grayscale
background in the plot (a) is the normalized potential
$\langle\psi\rangle=[e/(mc^2)]\langle A_z-\Phi\rangle$, where
$\langle\cdots\rangle$ means the averaging over the laser period
$2\pi/\omega_0$, $\langle A_z\rangle$ and $\langle\Phi\rangle$ are
vector and scalar potentials associated with low-frequency
wakefields. This potential determines the low-frequency electric
and magnetic fields in plasma, and, hence, the forces acting on
test electrons. The plot (a) shows that accelerated particles are
collected in the focusing and accelerating quarter of the
wakefield period. Quite a few electrons stay near the bottom of
the potential well with the energy gain, as is testified by the
plot~(c), almost negligible. The transverse spread of the bunch is
reduced roughly twice if compared with the initial value.

\begin{figure}[t]
\includegraphics[scale=1]{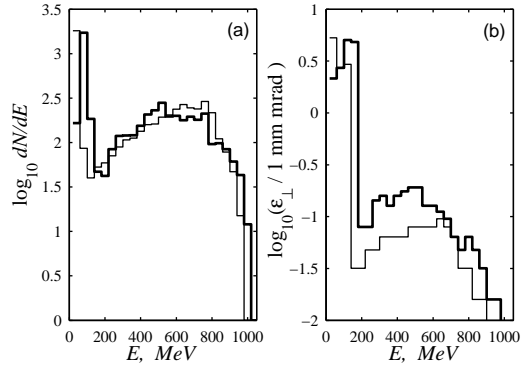}
\caption{\label{Fig6}Energy spectrum (a) and emittance (b) at the
extraction plane $z=Z_R$  for electron bunches  injected with
$\gamma_{e0}=125$ (thick lines) and $\gamma_{e0}=42$  (thin lines)
into the second wake period.
   }
\end{figure}

Figure~\ref{Fig6} shows the final energy and emittance of the test
electrons collected by equally distributed channels of an electron
spectrometer.  We compare them for the two different resonant
condition fulfilled at the injection plane. The thick lines
correspond to $\gamma_{e0}=\gamma_g=125$ (resonance with the laser
pulse) and the thin ones to $\gamma_{e0}=\gamma_{p0}\approx42$.
This value of $\gamma_{e0}$ corresponds to the resonance with the
second period of the plasma wake when the injection point is on
the axis. In both cases the energy spectrum looks as a shoulder
with not too pronounced maxima near 0.5~GeV for $\gamma_{e0}=125$
and 0.75~GeV for $\gamma_{e0}=42$. There is no explicit energy gap
separating the accelerated electrons from those which are
non-accelerated. In the case of $\gamma_{e0}=42$ the final
emittance is smaller than that for $\gamma_{e0}=125$ [see the plot
(b)]. The energy cutoff appears to be independent of the injected
electron energy and is close to 1~GeV.

The energy limit of 1 GeV can be exceeded if the electrons are
injected earlier and extracted from the wake later. Additional
runs show that elongating the acceleration length by 50\%
($-1.5Z_R\le z\le1.5Z_R$) gives a 15\% increase in the maximum of
final energy. As the highest acceleration gradients are achieved
near the laser focus, increasing the acceleration length beyond
$2Z_R$ does not lead to a substantial growth of electron energy.

Analysis of Figs.~\ref{Fig5} and~\ref{Fig6} leads to the following
conclusion: injection of a quite long (of the order of or longer than a wake
period) resonant electron bunch results in a large energy
spread of trapped and accelerated electrons no matter whether the
injected bunch was resonant with the laser pulse or with a given
period of the wake. It is obvious that the resonant conditions
$\gamma_{e0}=\gamma_g$ or $\gamma_{e0}=\gamma_{p0}$ serve well
only for the electrons injected in just one quarter (focusing and
accelerating) of the wake period. Sample simulation made for a
short electron bunch ($11.5\le k_p\xi \le 12.5$) with
$\gamma_{e0}=125$ and zero initial emittance gave the mean output
0.85 GeV per electron with about 30\% energy spread.

\section{\label{Sec5}Acceleration of non-resonant electron bunches}

\begin{figure}[t]
\includegraphics[scale=1]{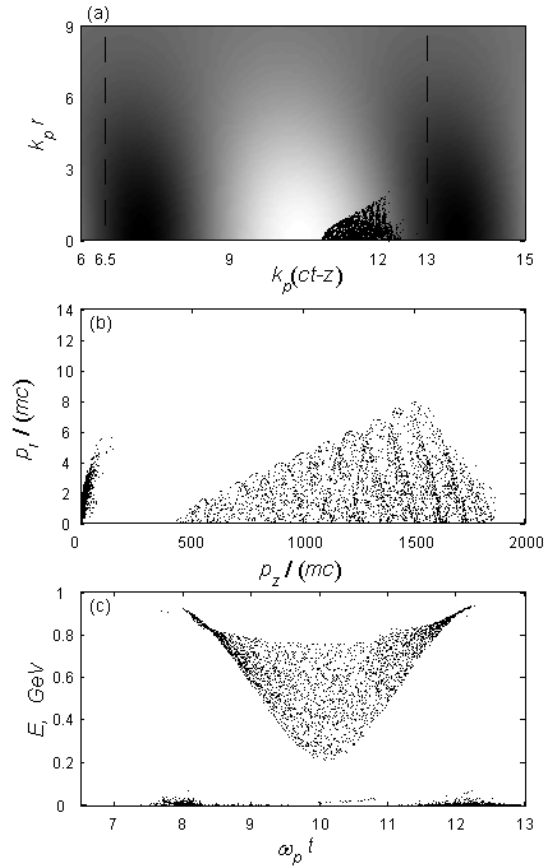}
\caption{\label{Fig7}Acceleration of the non-resonant electron
bunch. All parameters being the same as in Fig.~\ref{Fig5}, except
of initial relativisitc factor of electrons
$\gamma_{e0}=10\ll\gamma_g$.
   }
\end{figure}

Now studying non-resonant injected electron bunches, we bear in
mind that the initial energy of electrons is essentially smaller
than that of the resonant ones. Figure~\ref{Fig7} presents the
same plots as Fig.5 but for injected electrons with
$\gamma_{e0}=10\ll\gamma_{g(p0)}$.

Comparison of Figs.~\ref{Fig5} and~\ref{Fig7} reveals certain
benefits of using the electrons non-resonant at injection. First,
the accelerated bunch of initially non-resonant particles is more
compact in the radial direction [compare plots (a) of the two
figures], and the radial momentum spread is lower [compare plots
(b)]. Fig.~\ref{Fig7}(a) shows no electrons at the bottom of the
potential well, $k_p\xi\approx10$, hence, all the trapped
non-resonant electrons are accelerated,  and the energy gap
finally appears between the trapped and non-trapped particles [see
Fig.~\ref{Fig7}(c)]. Opposite to the resonant case, the
non-resonant electrons, loaded into either accelerating or
decelerating focusing phases, gain almost the same energy. At the
point of injection, the wake phase outruns the non-resonant
particles, so they slip out of the disadvantageous phase towards
the focusing and accelerating one. Otherwise, the resonant
electrons are initially at rest in the decelerating phase, and
their slippage takes more time. So, their final energy proves to
be lower.

\begin{figure}[t]
\includegraphics[scale=1]{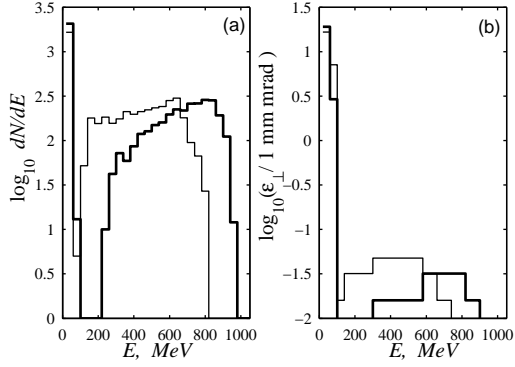}
\caption{\label{Fig8}Final energy spectrum (a) and emittance (b)
of electrons injected in different wake periods. The electron are
injected with $\gamma_{e0}=10$. Thick and thin lines correspond to
the injection to the second (data from simulation of
Fig.~\ref{Fig7}) and sixth period, respectively.
   }
\end{figure}

Figure~\ref{Fig8} shows final energy spectrum and emittance of
initially non-resonant test electrons with $\gamma_{e0}=10$,
injected at $z_0=-Z_R$ with zero initial emittance in the second
($6.5\le k_p\xi\le13$, simulation of Fig.~\ref{Fig7}) and sixth
($32.1 \le k_p\xi \le38.4$) wake periods. The initial rms radius
of the bunch is the same as in Figs.~\ref{Fig5} and~\ref{Fig7}.
The electrons accelerated in the second wake period are separated
from the bulk of low energetic particles by a 150~MeV wide gap
[see also Fig.~\ref{Fig7}(c)]. Besides, efficiency of acceleration
reduces when  the time delay between the laser pulse and electron
bunch grows.  The spectrum of electrons accelerated in the second
period has the maximum at 0.8~GeV and a sharp cutoff at 0.95~GeV.
Acceleration in the sixth period gives a plateau-like energy
spectrum which rises steadily up to 0.65~GeV and drops at
0.85~GeV. Therefore, the injection time lag should be taken as
small as possible in order to reduce the adverse effect of the
wake phase velocity variation. As compared with the resonant case
[Fig.~\ref{Fig6}(b)], final emittance of the bunch  is typically
lower in the non-resonant case (by a factor of three for the
spectrometer bins beyond 500 MeV).
\begin{figure}[t]
\includegraphics[scale=1]{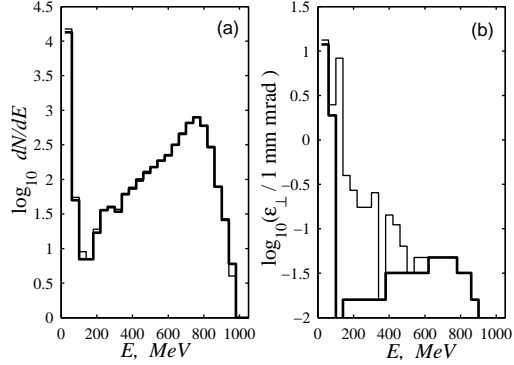}
\caption{\label{Fig9} Energy spectrum (a) and rms emittance (b) of
accelerated electrons for non-resonant ($\gamma_{e0}=10$)  wide
($k_p\sigma=13.5$) and long ($6.5\le k_p\xi \le25.4$) electron
bunch with zero initial emittance (thick lines) and with an
initial rms emittance $\varepsilon_\perp=2.145$~mm~mrad (thin
lines).
   }
\end{figure}

Technological limitations of monoenergetic electron injectors lead
inevitably to using in the experiment long ($\tau_b\gg\tau_p$) and
wide ($k_p\sigma\gg2\pi$) electron bunches with nonzero rms
emittance. The results of sample modelling of this situation are
presented in Fig.~\ref{Fig9}. The energy spectrum (a) and rms
emittance (b) of accelerated electrons are shown at the extraction
point $z=Z_R$. The non-resonant ($\gamma_{e0}=10$) electron beam
injected at $z_0=-Z_R$ covers three consecutive wake periods,
$6.5\le k_p\xi \le25.4$. The transverse size of the beam
$k_p\sigma=13.5$ is by a factor of three larger than that of
Figs.~\ref{Fig5} and~\ref{Fig7}, and the number of particles in
the bunch is 20000. The thick lines correspond to the bunch with
zero rms emittance at injection, and the thin lines to the case
with initial rms emittance $\varepsilon_\perp=2.145$~mm~mrad.
Variation of the initial rms emittance has a negligible effect on
the energy spectrum which has a sharp maximum at 0.75~GeV and
cutoff at 0.95~GeV. Only emittance of low-energy electrons
(spectrometer bins at $E<500$~MeV) is increased due to nonzero rms
emittance of the injected beam. The emittance of the higher-energy
electrons is unaffected. Thus, the simulations show that
elongating the electron bunch still preserves all the benefits of
using the non-resonant electrons and does not create any challenge
for the experimental implementation of the proposed scheme.

\begin{figure}[t]
\includegraphics[scale=1]{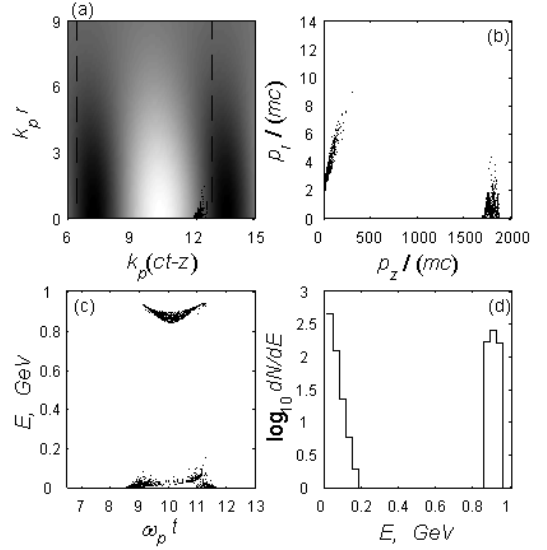}
\caption{\label{Fig10}Acceleration of low-energy electron bunch
($\gamma_{e0}=2$). Plots (a)-(c) show the same quantities as in
Figs.~\ref{Fig5} and~\ref{Fig7}. The electron energy spectrum is
shown in plot (d). Injection of highly non-resonant bunch provides
almost monoenergetic acceleration of trapped electrons.
   }
\end{figure}
Effective trapping and acceleration of electrons also occurs for
bunches injected with very low energy ($E_{in}\sim1$~MeV), which
lead to almost monoenergetic acceleration~\cite{16}. In
Fig.~\ref{Fig10}, the energy spectrum of accelerated electrons
with initial relativistic factor $\gamma_{e0}=2$ and radial spread
$k_p\sigma=2$ is shown while the other parameters are taken the
same as of Figs.~\ref{Fig5} and~\ref{Fig7}. The total number of
injected particles is 2500, of which roughly one half is trapped
and accelerated up to 900~MeV with 10\% energy spread. The final
rms emittance of accelerated bunch is lower than
$10^{-4}$~mm~mrad.

\section{\label{Sec6}Conclusion}

Construction of the petawatt ultra-short-pulse lasers will create
an opportunity for standard LWFA of electrons up to GeV energies.
In quite tenuous plasmas ($\omega_p<\tau_L^{-1}$) the relativistic
and ponderomotive nonlinearities of a not very tightly focused
($k_pr_0\ge2\pi$) laser pulse cancel each other~\cite{8}. In this
regime, an overcritical laser pulse propagates like in vacuum.
Increasing the laser focal spot up to 100 $\mu$m in radius
elongates the laser-plasma interaction length (estimated as twice
the Rayleigh length) up to 8 cm without any external optical
guiding. Such not very tightly focused laser pulse, however, has
enough energy to drive a nonlinear plasma wakefield along this
distance, which provides a controllable acceleration of externally
injected electrons up to 1 GeV. The nonlinear features of
quasi-plane wake plasma wave facilitate trapping, focusing, and
acceleration of the electrons from the injected bunch. The
converging laser pulse drives the plasma wake whose period
stretches (due to relativistic decrease in plasma frequency) as
the pulse approaches the focal plane. At this stage, the wake
phase velocity drops below the pulse group velocity. Growth of the
wake amplitude combined with the decrease in the phase velocity
provides the efficient trapping of low-energy (non-resonant)
electrons. Beyond the focal plane, the laser pulse diverges, and
the wake phase may become superluminous, which makes for
additional longitudinal compression of the electron bunch in the
focusing and accelerating quarter of the wakefield period. These
features of wakefield structure reduce the energy spread and
emittance of the electron bunch injected out of resonance with the
wake, $\gamma_{e0}\ll\gamma_{g(p0)}$, and, in the limit of very
low initial energy (1 MeV), provide almost monoenergetic
acceleration.

\begin{acknowledgments}
L. M. G. and S. Yu.\ K. gratefully acknowledge the hospitality of
CPhT, \'Ecole Polytechnique. This work was partly supported by
Centre National de la Recherche Scientifique (France) in the frame
of cooperation with the Russian Academy of Sciences, the \'Ecole
Polytechnique (in the form of postdoctoral fellowship for S. Yu.\
K.), the Russian Fund for Basic Research (Grant No. 02-02-1611),
and by the U.S. Dept. of Energy under Contracts No.
DE-FG02-04ER54763 and DE-FG02-04ER41321.
\end{acknowledgments}

\appendix

\section{\label{AppA}Nonlinear effects in propagation of ultra-short
laser pulses}

Self-consistent evolution of axi-symmetric laser pulse and
perturbations of electron plasma density can be described in the
weakly nonlinear quasi-static approximation by the set of
equations~\cite{17},
\begin{subequations}
\label{A1}
\begin{eqnarray}
2ik_0\frac{\partial a}{\partial z} +\Delta_\perp a & = &
k_p^2 \left(N-\frac{1}{4}|a|^2\right)a,\label{A1a}\\
\frac{\partial^2 N}{\partial \xi^2}+k_p^2N & = &\frac14
\left(\frac{\partial^2 }{\partial \xi^2}
   + \Delta_\perp\right)|a|^2,\label{A1b}
\end{eqnarray}
\end{subequations}
where $a=eE/(m\omega_0c)$ is the normalized amplitude of the laser
electric field,  $N\equiv (n_e-n_0)/n_0$ is the normalized
electron density perturbation, and $\Delta_\perp\equiv
r^{-1}\partial/\partial r(r\partial/\partial r)$ is the radial
part of the Laplace operator.

In the limit of a short and wide pulse whose length $c\tau_L$ is
smaller but the radius  $r_0$ is larger than $k_p^{-1}$, the main
terms in Eq.~(\ref{A1b}) are those containing the second
derivatives with respect to $\xi$. In this approximation, the
nonlinear terms in the right-hand side (RHS) of Eq.~(\ref{A1a})
cancel each other, and the pulse propagates as in vacuum~\cite{8}.
The small corrections to the index of refraction which are due to
the finite pulse length and proportional to $(\omega_p\tau_L)^2$
were considered in Ref.~\onlinecite{8}. We consider the effect of
small longitudinal and transverse nonlinearities on the evolution
of a short laser pulse and assume the transverse pulse shape be
Gaussian in every cross-section,
\begin{equation}
\label{A2} a \! = \!
\frac{A_0(\xi)}{f(\xi,z)}\exp\left[-\frac{r^2}{r_0^2f^2} \! + i\!
\left(\frac{r^2}{2}\beta(\xi,z) \! + \!
\varphi(\xi,z)\right)\right].
\end{equation}
Here,  $f(\xi,z)$ is the dimensionless focal spot size (pulse
width) which equals unity in the focal plane $z=0$ [i.e.,
$f(\xi,0)\equiv1$]. The function $A_0(\xi)$ gives an initial
amplitude profile on axis, the initial focal spot radius is $r_0$,
the quantities $\varphi(\xi,z)$ and $\beta(\xi,z)$ give the
on-axis values of phase and curvature of the laser phase front,
respectively. Substituting Eq.~(\ref{A2}) into Eqs.~(\ref{A1})
gives the equation for the pulse width~\cite{18},
\begin{eqnarray}
\lefteqn{ \frac{\partial^2 f}{\partial z^2} \! - \!
\frac{f^{-3}}{Z_R^2} \! = \! -\frac{f}{4}
\frac{\omega_pk_p^3}{\omega_0Z_R}}\label{A3}\\ & & \times
{}\int_{-\infty}^{\xi}d\xi'\sin[k_p(\xi \! - \! \xi')]
\frac{A_0^2(\xi')}{f^4(\xi',z)}\left(1\! + \!
\frac{16}{k_p^2r_0^2f^2(\xi',z)}\right),\nonumber
\end{eqnarray}
where $z$ is the coordinate of the pulse center moving from the
left to the right; the transverse Laplace operator in
Eq.~(\ref{A1b}) gives the second term in the brackets in the
integrand. In the short-pulse case, $A_0^2(\xi)$ is nonzero within
an interval  $|\xi|\ll k_p^{-1}$. Integrating by parts the
expression in the RHS of Eq.~(\ref{A3}) and taking account of only
the main term gives
\begin{eqnarray}
   \lefteqn{\frac{\partial^2 f}{\partial z^2} \! - \!
\frac{f^{-3}}{Z_R^2} \! = \! - \frac{f}{4}
\frac{\omega_pk_p^3}{\omega_0Z_R}}\label{A4}\\
& & {} \times \! \int_{-\infty}^{\xi}d\xi'\int_{-\infty}^{\xi'} \!
d\xi'' \frac{A_0^2(\xi'')}{f^4(\xi'',z)}\left(1 \! + \!
\frac{16}{k_p^2r_0^2f^2(\xi'',z)}\right).\nonumber
\end{eqnarray}
The laser pulse of small amplitude propagates in plasma as in
vacuum. In order to evaluate the threshold amplitude $a_{0c}$
above which the effect of nonlinearities might occur, we consider
Eq.~(\ref{A4}) near the focal plane. At the focal plane $z=0$ the
pulse width $f$ is constant while $A_0(\xi)$ alters within a
relatively short interval $|\xi|\ll k_p^{-1}$. We assume that
$f(\xi,z)$ varies with $\xi$ slower than $A_0(\xi)$ within some
segment of the laser path. Equation~(\ref{A4}) then reduces to
\begin{equation}
\label{A5} \frac{\partial^2 f}{\partial \zeta^2} -
\frac{1}{f^3}\left\{1-\alpha I(\xi)\left(1+\frac{2}{\alpha
f^2}\right)\right\}=0,
\end{equation}
where $\zeta=z/Z_R$ is the normalized propagation distance,
$\alpha=(k_pr_0)^2/8$ is the  normalized initial squared spot size
\begin{equation}
\label{A6} I(\xi) = k_p^2\int_{-\infty}^\xi d\xi'
\int_{-\infty}^{\xi'}d\xi'' A_0^2(\xi'').
\end{equation}
The condition $f(\xi,0)\equiv1$ gives the integral of
Eq.~(\ref{A5}),
\begin{equation}
\label{A7} \left(\frac{\partial f}{\partial \zeta}\right)^2 -
\left(1-\frac{1}{f^2}\right)\left[1-\alpha I-
I\left(1+\frac{1}{f^2}\right)\right]=0.
\end{equation}
Initially, the evolution of a small spot size perturbation $\delta
f=f-1$, $|\delta f|<1$, obey the equation following from
Eq.~(\ref{A7}),
\begin{equation} \label{A8}
(\partial \delta f/\partial q)^2 - 2\delta f(B-1)=0,
\end{equation}
where $q=\zeta\sqrt{I}$, and $B=I^{-1}-1-\alpha$.
Equation~(\ref{A8}) has real solutions for both $B<1$ (then
$\delta f<0$) and $B>1$ (then $\delta f>0$). Reduction of the
pulse width in the first case corresponds to the converging laser
pulse, while in the second case the laser diverges. The border of
laser pulse stability against the transverse distortions, $B=1$,
gives the implicit equation for the stability point,
\begin{equation}
\label{A9}I(\xi_c)=(2+\alpha)^{-1}.
\end{equation}
The point $\xi_c$ separates diverging and converging parts in the
pulse profile. The function $I(\xi)$  grows monotonically from the
leading front towards the trailing edge of the pulse. Therefore,
the pulse portion which lays between the leading front and the
point $\xi_c$ spreads, while the part beyond $\xi_c$ focuses. As a
result, in the process of propagation the short pulse acquires the
form similar to "beet-root". In the vicinity of the threshold
point $\xi_c$, the pulse width given by  Eq.~(\ref{A8}) reads
$f=1-(\zeta^2/2)[I(\xi)-I(\xi_c)]/I(\xi_c)$.

We examine the case of a Gaussian laser pulse,
$A_0(\xi)=a_0\exp[-2\ln2\xi^2/(c\tau_L)^2]$, with the center
located at $\xi=0$. Substituting $A_0(\xi)$ into Eq.~(\ref{A6})
gives
\begin{equation}
\label{A10} I(\xi) \! = \!
\frac{(a_0\omega_p\tau_L)^2}{8\ln2}\int_{-\xi2\sqrt{\ln2}/(c\tau_L)}
^{\infty}dx\left[1-\Phi(x)\right],
\end {equation}
where
\[
  \Phi(x)=\frac{2}{\sqrt\pi}
\int_{0}^{x}dt\exp(-t^2)
\]
is the probability integral. Substantial modification of the laser
pulse shape occurs when the threshold condition (\ref{A10}) is met
at the pulse center (i.e., $\xi_c=0$). In consistence with
Eq.~(\ref{A10}), this criterion takes the form
\begin{equation}
\label{A12} a_{0c}=\frac{1}{\omega_p\tau_L}
\sqrt{\frac{4\ln2}{1+(k_pr_0/4)^2}}.
\end{equation}

\section{\label{AppB} The wakefield of an ultra-short focusing
laser pulse}

The wakefield is excited by the ponderomotive force associated
with a laser pulse. Under the assumption that the linear theory of
diffraction of Gaussian beams~\cite{12} holds, the laser pulse
envelope $a(r,z,\xi)$ reads
\begin{equation}
\label{B1} a \! = \! \frac{a_0}{\sqrt{1 \! + \!
\zeta^2}}\exp\left[-2\ln 2\frac{\xi^2}{(c\tau_L)^2}\! - \!
\frac{r^2}{r_0^2(1 \! + \! \zeta^2)} \! + \! i\Psi\right],
\end{equation}
where  $\zeta=z/Z_R$, $\xi=v_gt-(z+z_0)$,
$\Psi=(r/r_0)^2\zeta/(1+\zeta^2)-\arctan\zeta$,  $v_g$ is the
group velocity of a pulse, other notations are the same as in
Eq.~(\ref{1}). At the initial moment $t=0$, the pulse center
$\xi=0$ resides at $z=-z_0$. The pulse propagates towards positive
$z$.

In the linear approximation, the frequency of electron plasma
oscillation equals $\omega_p$,  and the phase velocity $v_p$ of
the wake coincides with the group velocity  $v_g$ of the pulse. In
the weakly nonlinear approximation, the frequency of plasma
oscillations is downshifted because of relativistic increase of
the mass of oscillating electron~\cite{19}. The intensity of the
Gaussian beam~(\ref{B1}) varies along the laser path, and so does
the amplitude of the plasma wake. Hence, relativistic reduction of
the plasma frequency alters as a function of propagation distance,
and this phase slippage characterizes the variation of the phase
velocity of the wake across the plasma. Consideration of this
effect is easier in the case of a wide, $k_pr_0\gg1$, and not so
intense, $a_0<1$, laser pulse when the wake electric field is the
potential one. Then, for the laser pulse amplitude taken in the
form (\ref{B1}), the dimensionless wake potential
$\phi=e\varphi/(mc^2)$ is given by~\cite{3}
\begin{eqnarray}
\lefteqn{\phi  =  -\frac{(a_0/2)^2g(x)}{1  +
\zeta^2}}\label{B2}\\&&\times \exp\left[-\frac{2r^2}{r_0^2(1  +
\zeta^2)}\right]\sin\left\{\left[k_p  + \frac{ \Delta
\omega_p(z,r)}{c}\right]\xi\right\},\nonumber
\end{eqnarray}
where $g(x)=x\sqrt{\pi/2}\exp(-x^2/8)$ depends on the
dimensionless pulse length $x=\omega_p\tau_L/\sqrt{2\ln2}$, the
relativistic frequency shift~\cite{19} is $ \Delta \omega_p =
-(3/16)\omega_p[(a_0/2)^2/(1 + \zeta^2)]^2 \exp[-(2r/r_0)^2/(1
+\zeta^2)]g^2(x)$. The on-axis phase of the wake potential follows
from Eq.~(\ref{B2}),
\begin{equation}
\label{B5} \theta =
k_p(v_gt-z)\left\{1-\frac{3}{16}\left[\frac{(a_0/2)^2g(x)}{1+\zeta^2}\right]^2\right\}.
\end{equation}
Then, the wake phase velocity $v_p$ reads
\begin{equation}
\label{B6} v_p \!= \!- \frac{\partial\theta/\partial
t}{\partial\theta/\partial z} \!\approx\! v_g \!\left\{1 \!+
\!\frac{3}{4} \! \left[\frac{(a_0/2)^2 g(x)}{1 +
\zeta^2}\right]^2\frac{\zeta\xi/Z_R}{1 + \zeta^2}\right\} \! .
\end{equation}
The quantity $\xi/v_g$ characterizes a positive time delay with
respect to the pulse center, and $\zeta$ characterizes the laser
pulse position relative to the focal plane. When the pulse passes
the focal plane, $\zeta$ becomes positive. The wake phase velocity
coincides with the pulse group velocity only far away from the focal
plane and exactly at the focal plane.  The wake phase velocity
decreases while the pulse approaches the focal plane, and reaches
the minimum value for $z=-Z_R/\sqrt{5}$ where $ v_p\approx
v_g\{1-0.192[(a_0/2)^2g(x)]^2\xi/Z_R\}$. The wake phase velocity
exceeds $v_g$ and can exceed the vacuum speed of light while the
pulse moves away from the focal plane and radially spreads. Small
variation of the phase velocity $v_p$ with respect to $v_g$ can,
however, considerably modify the Lorentz factor
$\gamma_{p0}\equiv(1-v_p^2/c^2)^{-1/2}$ if compared with
$\gamma_g\equiv(1-v_g^2/c^2)^{-1/2}$. Eq.~(\ref{B6}) gives
\begin{equation}
\label{B8} \gamma_{p0}^{-2}  \approx  \gamma_{g}^{-2}  + \frac32
\left[\frac{(a_0/2)^2g(x)}{1+\zeta^2}\right]^2\frac{\zeta\xi/Z_R}{1+\zeta^2}.
\end{equation}
The electrons are injected at negative $\zeta$ where $v_p<v_g$.
Hence, the resonance with the nonlinear wakefield needs less
electron energy ($mc^2\gamma_{p0}$) than the resonance with the
laser pulse (or a linear wakefield), $mc^2\gamma_g$.

Although Eq.~(\ref{B8}) is weakly relativistic, we use it for a
qualitative estimate of resonant gamma-factors for different wake
periods for the simulation  parameters of Sec.~\ref{Sec4}
and~\ref{Sec5} ($\omega_p\tau_L=0.56$, $a_0=1.72$, $k_pr_0=6.3$,
$Z_R=4$~cm, $\gamma_g=125$). The accelerating phase of the second
period (See Fig.~\ref{Fig5}) corresponds to $k_p\xi\approx6.5$
(i.e., $\xi=-0.01$~cm), and, at the point of injection $\zeta=-1$,
has a gamma-factor $\gamma_{p0}\approx80$. For the accelerating
phase of the fifth plasma period ($k_p\xi\approx25$)
Eq.~(\ref{B8}) gives $\gamma_{p0}\approx49$. Graphic estimates of
resonant Lorentz factors in Fig.~\ref{Fig4}(b) exceed the weakly
relativistic estimates roughly by a factor two.














\end{document}